# THE TRACE IDENTITY AND THE PLANAR CASIMIR EFFECT

S. G. Kamath*

Department of Mathematics, Indian Institute of Technology Madras,

Chennai 600 036,India

Abstract:

The familiar trace identity associated with the scale transformation $x^\mu \to x'^\mu = e^{-\lambda} x^\mu$ on the Lagrangian density for a noninteracting massive real scalar field in 2 + 1 dimensions is shown to be violated on a single plate on which the Dirichlet boundary condition $\varphi(t, x^1, x^2 = -a) = 0$ is imposed. It is however respected in: i. 1 + 1 dimensions in both free space and on a single plate on which the Dirichlet boundary condition $\varphi(t, x^1 = -a) = 0$ holds; and, ii. in 2 + 1 dimensions in free space, i.e. the unconstrained configuration. On the plate where $\varphi(t, x^1, x^2 = -a) = 0$, the modified trace identity is shown to be *anomalous* with a numerical coefficient for the anomalous term equal to the canonical scale dimension viz. ½. The technique of Bordag, Robaschik and Wieczorek [5] is used to incorporate the said boundary condition into the generating functional for the connected Green's functions.

# 1. Introduction

The Casimir energy [1] of quantized fields has been conventionally defined as a difference in zero-point energy. In other words, if $E_0[\partial\Gamma]$ represents the zero-point energy in the presence of boundaries and $E_0[0]$ that without boundaries, then the Casimir energy is formally defined as

$$E_{vac}[\partial\Gamma] = E_0[\partial\Gamma] - E_0[0] \tag{1}$$

In terms of the energy – momentum tensor $\hat{\Theta}^{\mu\nu}$ (1) rewrites as [2] the volume integral of

$$\Theta^{\mu\nu}_{vac}[\partial\Gamma] = \langle 0|\hat{\Theta}^{\mu\nu}|0\rangle_{\partial\Gamma} - \langle 0|\hat{\Theta}^{\mu\nu}|0\rangle_0 \tag{2}$$

with $\mu = \nu = 0$. Thus the measurable energy density of the vacuum would be defined as the difference between that in the constrained field configuration and the one corresponding to the unconstrained field. Of especial relevance to this paper is the fact that $\Theta^{\mu\nu}_{vac}[\partial\Gamma]$ can be expressed in terms of the field propagators.

To take this connection further and to elaborate further below on the motivation of this paper we recall that the trace identity associated with the massless free real scalar field Lagrangian density in 3 + 1 dimensions, $L = \frac{1}{2}\partial^\nu\phi\,\partial_\nu\phi$, is given by [3]

$$g_{\mu\nu}\Gamma^{\mu\nu}(y;x,z) + id\left(\delta^{(4)}(y-x) + \delta^{(4)}(y-z)\right)G(x,z) = 0 \tag{3}$$

with the canonical scale dimension $d = 1$; $G(x, z)$ and $\Gamma^{\mu\nu}(y;x,z)$ being the connected T*-ordered products $\langle 0|T^*(\phi(x)\phi(z))|0\rangle_c$ and $\langle 0|T^*(\hat{\Theta}^{\mu\nu}(y)\phi(x)\phi(z))|0\rangle_c$ respectively. To get our point across below we have temporarily chosen to side-step the formal derivation of eq.(3); a somewhat more detailed discussion using the methods of Coleman and Jackiw [3] will be presented later in the paper.

Let us now motivate the work reported here: Labelling the two terms in eq.(3) by I and J respectively it is now natural to ask if in the light of $\Theta_{vac}^{\mu\nu}[\partial\Gamma]$ being non-zero, would

$$I_{vac}[\partial\Gamma] = I[\partial\Gamma] - I[0] \tag{4a}$$

also be non-zero? By (3), this difference should now be the negative of

$$J_{vac}[\partial\Gamma] = J[\partial\Gamma] - J[0] \tag{4b}$$

with the first and second terms on the r.h.s. of eqs.(4) being the corresponding Green's functions in the constrained and unconstrained field configurations respectively, the Dirac $\delta$ - function being included under the label J. With the Lagrangian for the noninteracting massive real scalar field given in eq.(6) below as our model we report herein that the trace identity

$$g_{\mu\nu}\Gamma^{\mu\nu}(y;x,z) + id\{\delta^{(3)}(y-x) + \delta^{(3)}(y-z)\}G(x,z) = \Delta(y:x,z) \tag{5}$$

is not preserved in 2 + 1 dimensions (wherein d = ½ ); or, to make our point, while (5) holds in the unconstrained configuration, it does not in the constrained version – see eq.(54) below - the latter differing from the former by the introduction of a single plate on which the Dirichlet boundary condition $\phi$ (t , $x^1$, $x^2$ = - a) = 0 is imposed. In contrast however, with the same model Lagrangian as in (6) below but in 1 + 1 dimensions wherein d = 0, the corresponding version of (5), namely

$$g_{\mu\nu}\Gamma^{\mu\nu}(y;x,z) = \Delta(y;x,z) \tag{5a}$$

is preserved in both the unconstrained and constrained field configurations. The latter being, again, distinguished by the introduction of a single plate on which the boundary condition $\phi$( t , $x^1$ = - a) = 0 is imposed. The r.h.s. of eqs. (5) and (5a) represent the connected Green's function $m^2 \langle 0|T^*(\varphi^2(y)\phi(x)\phi(z))|0\rangle_c$, m being the mass of the scalar field in (6).

To stress our point of view therefore, this paper shows that within the framework of the Casimir effect there is a qualitative difference between 1 + 1 and 2 + 1 dimensions in so far as the maintenance of the trace identity represented by eq.(5) is concerned. The difference manifests, as will be evident later, in the appearance of an *anomalous* term with a numerical coefficient equal to the canonical scale dimension d = ½ in 2 + 1 dimensions – see eq.(60) and the succeeding remarks below - there being none in 1 + 1 dimensions as d = 0 in this case.

The introduction of a single plate on which the Dirichlet boundary condition holds has been done here to focus on not the Casimir effect per se, it being now well established experimentally [4], but to underline as a byproduct of this report how even a minimal change in the unconstrained configuration, i.e. free space, implied by the presence of a single plate as a boundary in the constrained case leads to hitherto unsuspected differences in 1 + 1 versus 2 + 1 dimensions in so far as the validity of the trace identity is concerned; indeed, the trace identity in 1 + 1 dimensions given by eq.(5a) will be found below to be trivially satisfied in both the configurations.

This *departure* from the use of a pair of parallel conducting plates separated by a distance that was adopted in the original Casimir calculation [1] to a single plate in this paper has been done deliberately and to the best of our knowledge the subject of this paper has not received attention in the published literature [1, 2 ,4 ] . We have however *persisted* with the use of the term 'Casimir effect' in the title of this paper if only to:

a. acknowledge that the work done here has been inspired by it and,
b. simultaneously highlight how even a minimal modification of the unconstrained configuration with a single plate as opposed to the *not so* minimal two plate version adopted in the Casimir calculation[1], leads to the conclusions reported here, deferring at the same time a recalculation of the Casimir energy density $\Theta_{vac}^{00}[\partial\Gamma]$ defined by eq.(2) for the single plate configuration to a subsequent report ; in doing so we anticipate that $\Theta_{vac}^{00}[\partial\Gamma]$ will be different from zero.

Put differently, it is the *susceptibility* if any, of the trace identity given by eqs.(5) and (5a) to the minimal modification of free space in the manner mentioned above that is on test and in focus in this paper; the evident difference between the results in 1 + 1 and 2 + 1 dimensions is only incidental.

The plan of this paper is as follows: In sec.2 we shall present following Coleman and Jackiw[3] a brief derivation of the trace identity associated with the scale transformation $x^\mu \to x'^\mu = e^{-\rho}x^\mu$ in both 1 + 1 and 2 + 1 dimensions; as stressed by these authors [3], a redefinition of the canonical energy momentum tensor $\hat{\Theta}_c^{\mu\nu}$ is done in this section particularly with respect to 2 + 1 dimensions to obtain a modified dilatation current without the field virial. Sec. 2 is also routine for the learned reader and is included here merely for the sake of completeness, with the explicit verification of the trace identities given in eqs.(5) and (5a) for the unconstrained configuration being taken up in sec.3. Sec.4 is the main part of this paper and we present therein firstly, an explicit verification of (5a) with a redefined generating functional - following Bordag, Robaschik and Wieczorek [5] - for the connected Green's functions which incorporates the Dirichlet boundary condition $\varphi(t, x^1 = -a) = 0$; this approach is then repeated for a similar check on the maintenance of eq.(5) in 2 + 1 dimensions. The paper concludes with a short Appendix that includes some of the relevant equations that one needs to work with in sec.4.

## 2. The derivation of the trace identity

We begin with the Lagrangian density

$$L = \frac{1}{2}\partial^\mu \varphi \, \partial_\mu \phi - \frac{1}{2} m^2 \phi^2 \tag{6}$$

for which the canonical energy momentum tensor is

$$\hat{\Theta}_c^{\mu\nu} = \partial^\mu \phi \partial^\nu \phi - g^{\mu\nu} L \tag{6a}$$

and the canonical dilatation current $D^\lambda$:

$$D^\lambda = x_\mu \hat{\Theta}_c^{\lambda\mu} \tag{7a}$$

$$D^\lambda = x_\mu \hat{\Theta}_c^{\lambda\mu} + \frac{1}{4}\partial^\lambda \phi^2 \tag{7b}$$

in 1 + 1 and 2 + 1 dimensions respectively, the second term in (7b) being the field virial [3], it being zero in 1 + 1 dimensions. For both versions of $D^\lambda$ above one has

$$\partial^\mu D_\mu = m^2 \phi^2 \equiv \Delta \tag{7c}$$

as the Lagrangian (6) is not invariant under the scale transformation $x^\mu \to x'^\mu = e^{-\rho} x^\mu$, the noninvariant term in (7c) being labelled $\Delta$.

With the definition $\sigma^{\mu\nu} = \frac{1}{4} g^{\mu\nu} \phi^2$, it is now simple to use eq.(A.17) of Ref. 3 and construct from eq.(A.19) there the tensor

$$X^{\lambda\rho\mu\nu} = \frac{1}{4}\left(g^{\mu\nu} g^{\lambda\rho} - g^{\lambda\nu} g^{\rho\mu}\right)\phi^2 \tag{8}$$

with $\sigma^\mu_{+\mu} = \frac{3}{4}\phi^2$. Note that $X^{\lambda\rho\mu\nu} = -X^{\mu\rho\lambda\nu} = -X^{\lambda\nu\mu\rho}$ ; as emphasized by Coleman and Jackiw [3] the utility of (8) lies in providing a simplified version of $D^\lambda$ in eq.(12) below. Indeed, with eq.(8) we can now define

$$\hat{\Theta}^{\mu\nu} = \hat{\Theta}^{\mu\nu}_c + \frac{1}{2}\partial_\lambda \partial_\rho \left(X^{\lambda\rho\mu\nu}\right) \tag{9}$$

so that $\partial_\mu \hat{\Theta}^{\mu\nu} = \partial_\mu \hat{\Theta}^{\mu\nu}_c$; additionally,

$$\frac{1}{2}\partial_\lambda \partial_\rho \left(X^{\lambda\rho\mu\nu} x_\nu\right) = \frac{1}{2} x_\nu \partial_\lambda \partial_\rho X^{\lambda\rho\mu\nu} - \frac{1}{4}\partial^\mu \phi^2 \tag{10}$$

With (9) and (10) eq. (7b) now becomes

$$D^\mu = x_\nu \hat{\Theta}^{\mu\nu} - \frac{1}{2}\partial_\lambda \partial_\rho \left(X^{\lambda\rho\mu\nu} x_\nu\right) \tag{11}$$

The antisymmetry property of the $X^{\lambda\rho\mu\nu}$ mentioned above now helps us to infer [3] that: a. the second term in (11) can be dropped and that b. $\partial_\mu D^\mu_M = \hat{\Theta}^\mu_\mu$ where

$$D^\mu_M = x_\nu \hat{\Theta}^{\mu\nu} \tag{12}$$

We shall now present below following Appendix B of Ref. 3, a short derivation of the trace identities using the Ward identities for the energy momentum tensor $\hat{\Theta}^{\mu\nu}_c$; the task being easier for the 1 + 1 dimensional case we shall take it up first. The Ward identity for the $\hat{\Theta}_c^{\mu\nu}$ used in (7a) is given by [3]

$$\partial_\mu^{(y)} \langle 0|T^*\left(\hat{\Theta}_c^{\mu\nu}(y)\phi(x)\phi(z)\right)|0\rangle_c + i\delta^{(2)}(y-x)\langle 0|T^*\left(\partial^\nu\phi(x)\phi(z)\right)|0\rangle_c$$
$$+ i\delta^{(2)}(y-z)\langle 0|T^*\left(\phi(x)\partial^\nu\phi(z)\right)|0\rangle_c = 0 \quad (13)$$

with that for the dilatation current being

$$\partial_\mu^{(y)} \langle 0|T^*\left(D^\mu(y)\phi(x)\phi(z)\right)|0\rangle_c + i\delta^{(2)}(y-x)\langle 0|T^*\left(\delta\phi(x)\phi(z)\right)|0\rangle_c$$
$$+ i\delta^{(2)}(y-z)\langle 0|T^*\left(\phi(x)\delta\phi(z)\right)|0\rangle_c = \langle 0|T^*\left(\Delta(y)\phi(x)\phi(z)\right)|0\rangle_c \quad (14)$$

where $\delta\phi(x) = x \cdot \partial \phi(x)$. With (7a) and (13), (14) reworks to

$$g_{\mu\nu}\langle 0|T^*\left(\hat{\Theta}_c^{\mu\nu}(y)\phi(x)\phi(z)\right)|0\rangle_c = \langle 0|T^*\left(\Delta(y)\phi(x)\phi(z)\right)|0\rangle_c \quad (15)$$

which is (5a). To get eq.(5) on the other hand, one should remember that eq.(12) involves not the canonical energy momentum tensor $\hat{\Theta}_c^{\mu\nu}$ but $\hat{\Theta}^{\mu\nu}$ as given by eq.(9); however, since the second term in (9) is antisymmetric, (13) holds unchanged save that one should replace $\hat{\Theta}_c^{\mu\nu}$ by $\hat{\Theta}^{\mu\nu}$ and change the superscript 2 on the Dirac δ- function to 3 as is appropriate for 2 + 1 dimensions; the counterpart of (13) would thus be

$$\partial_\mu^{(y)} \langle 0|T^*\left(\hat{\Theta}^{\mu\nu}(y)\phi(x)\phi(z)\right)|0\rangle_c + i\delta^{(3)}(y-x)\langle 0|T^*\left(\partial^\nu\phi(x)\phi(z)\right)|0\rangle_c$$
$$+ i\delta^{(3)}(y-z)\langle 0|T^*\left(\phi(x)\partial^\nu\phi(z)\right)|0\rangle_c = 0 \quad (16)$$

Similarly, the Ward identity for the canonical dilatation current represented by eq.(14) will now be altered in the 2 + 1 dimensional case by the use of:

i. $D_M^{\ \mu}$ given by eq.(12) instead of $D^\mu$ and,

ii. $\delta\phi(x) = \left(\frac{1}{2} + x \cdot \partial\right)\phi(x)$, and lastly, the change of the superscript 2 to 3. Thus eq.(14) becomes

$$\partial_\mu^{(y)} \langle 0|T^*\left(D_M^\mu(y)\phi(x)\phi(z)\right)|0\rangle_c + i\delta^{(3)}(y-x)\langle 0|T^*\left(\delta\phi(x)\phi(z)\right)|0\rangle_c$$
$$+ i\delta^{(3)}(y-z)\langle 0|T^*\left(\phi(x)\delta\phi(z)\right)|0\rangle_c = \langle 0|T^*\left(\Delta(y)\phi(x)\phi(z)\right)|0\rangle_c \quad (17)$$

Using the same steps which led from (13) and (14) to (15) now yields the trace identity (5).

### 3. A check of eqs.(5) and (5a) without boundaries

We shall now take up for explicit verification the two trace identities for the *unconstrained* configuration starting with (15). For convenience we start with the r.h.s. of this equation and so let's consider the generating functional for the connected Green's functions given by

$$\exp(iZ[J,K]) = \frac{1}{W_0}\int[D\phi]\exp i\int d^2x\left(L + J\phi + Km^2\phi^2\right)$$

$$= \frac{1}{W_0}\int[D\phi]\sum_0^\infty \frac{(-i)^n}{n!}\left(\int d^2x\, m^2 K(x)\frac{\delta^2}{\delta J^2(x)}\right)^n \exp i\int d^2x(L+J\phi)$$

$$= \frac{1}{W_0}\sum_0^\infty \frac{(-i)^n}{n!}\left(\int d^2x\, m^2 K(x)\frac{\delta^2}{\delta J^2(x)}\right)^n \exp-\frac{1}{2}i\int d^2x d^2y J(x)\Delta(x-y)J(y) \quad (18)$$

with J and K being external sources, $W_0$ a constant so that the l.h.s. is unity when $J = K = 0$, and

$\Delta(x-y) = \int \frac{d^2p}{(2\pi)^2}\frac{\exp i\, p\cdot(x-y)}{p^2-m^2+i\varepsilon}$. For later use we note that:

$$(\partial_{(x)}^\mu \partial_\mu^{(x)} + m^2)\Delta(x-y) = -\delta^{(2)}(x-y) \quad (18a)$$

$$i\frac{\delta^2 Z[J,K]}{\delta J(s)\delta J(t)}\bigg|_{J=0=K} = i^2\langle 0|T^*[(\phi(s)\phi(t)]|0\rangle_c = -i\Delta(s-t) \quad (18b)$$

Eq.(18) yields

$$i\frac{\delta^3 Z[J,K]}{\delta K(u)\delta J(s)\delta J(t)}\bigg|_{J=0=K} \equiv i^3\langle 0|T^*\left(\Delta(u)\phi(s)\phi(t)\right)|0\rangle_c = -2im^2\left(-i\Delta(t-u)\cdot -i\Delta(s-u)\right)$$

so that by definition

$$\Delta(u;s,t) = 2m^2\left(-i\Delta(t-u)\right)\left(-i\Delta(s-u)\right) \tag{19}$$

For the l.h.s. of eq.(15) we use

$$\exp iZ\left[J, K^\alpha, D^{\mu\nu}\right] = \frac{1}{W_0}\int[D\phi]\exp i\int d^2x\left(L + J\phi + K^\alpha\partial_\alpha\phi + D_{\mu\nu}\Theta_c^{\mu\nu}\right)$$

$$= \frac{1}{W_0}\sum_0^\infty \frac{i^n}{n!}\left[\int d^2x D_{\mu\nu}(x)\hat{\Theta}_c^{\mu\nu}\right]^n \exp i\int d^2x\left(L + J\phi + K^\alpha\partial_\alpha\phi\right) \tag{20}$$

with $\hat{\Theta}_c^{\mu\nu}$ being given by (6a), and J(x), $K^\mu$(x) and $D^{\mu\nu}$(x) being external sources. It is now easy to rewrite (20) as

$$\exp iZ\left[J, K^\alpha, D^{\mu\nu}\right]$$
$$= \frac{1}{W_0}\sum_0^\infty \frac{(-i)^n}{n!}\left[\int d^2x D^{\mu\nu}H_{\mu\nu}\right]^n \exp -\frac{1}{2}i\int d^2x d^2y\left(J + K^\mu\vec{\partial}_\mu\right)_x \Delta(x-y)\left(J + \overleftarrow{\partial}_\mu K^\mu\right)_y \tag{21}$$

with

$$H_{\mu\nu} = \frac{\delta^2}{\delta K^\mu(u)\delta K^\nu(u)} - \frac{1}{2}g_{\mu\nu}\left(\frac{\delta^2}{\delta K^\alpha(u)\delta K_\alpha(u)} - m^2\frac{\delta^2}{\delta J(u)\delta J(u)}\right) \tag{21a}$$

Note that the derivative operators in the exponent in (21) act on the $\Delta$(x-y) defined earlier. Eq.(21) now leads to

$$i\frac{\delta^3 Z}{\delta D^{\alpha\beta}(u)\delta J(s)\delta J(t)}\bigg|_{J=0=K=D^{\alpha\beta}} \equiv i^3\langle 0|T^*\left(\hat{\Theta}_{\alpha\beta}(u)\phi(s)\phi(t)\right)|0\rangle$$

$$= -2i\left\{\left(-i\partial_\alpha^{(u)}\Delta(t-u)\right)\left(-i\partial_\beta^{(u)}\Delta(s-u)\right) - \frac{1}{2}g_{\alpha\beta}\left[\begin{array}{c}\left(-i\partial_\nu^{(u)}\Delta(t-u)\right)\left(-i\partial_{(u)}^\nu\Delta(s-u)\right) - \\ m^2\left(-i\Delta(t-u)\right)\left(-i\Delta(s-u)\right)\end{array}\right]\right\} \tag{22}$$

so that

$$g_{\alpha\beta} \langle 0|T^*\left(\hat{\Theta}_c^{\alpha\beta}(u)\phi(s)\phi(t)\right)|0\rangle = 2m^2\left(-i\Delta(t-u)\right)\left(-i\Delta(s-u)\right) \tag{23}$$

Identifying eq.(19) with the r.h.s. of (23) thus verifies (15), i.e. the trace identity in 1 + 1 dimensions.

We shall now verify eq.(5) below. It is enough to focus on the first term as the rest of the identity will follow naturally and to this end let's now consider

$$\exp iZ[J, K^\mu, D^{\alpha\beta}, L^{\mu\nu}] = \frac{1}{W_0}\int[D\phi]\ \exp i\int d^3x\left(L + J\phi + K^\nu\partial_\nu\phi + L^{\alpha\beta}\partial_\alpha\partial_\beta\phi + D_{\mu\nu}\hat{\Theta}_M^{\mu\nu}\right)$$

$$= \frac{1}{W_0}\sum_0^\infty \frac{(i)^n}{n!}\left\{\int d^3y D_{\mu\nu}(y)\hat{\Theta}_M^{\mu\nu}\right\}^n \exp-\left\{\frac{1}{2}i\int d^3x d^3y M(x)\Delta(x-y)M(y)\right\} \tag{24}$$

with the operator

$$M = J(x) + K^\alpha\partial_\alpha^{(x)} + L^{\mu\nu}\partial_\mu^{(x)}\partial_\nu^{(x)} \tag{24a}$$

the derivative operators in (24) acting on $\Delta(x-y)$, the latter being now defined by (18a) but with the $\delta$-function appropriate to 3 space-time dimensions. Rewriting (24) as

$$\exp iZ\left[J(x), K^\mu, D^{\alpha\beta}, L^{\mu\nu}\right] = \frac{1}{W_0}\sum_0^\infty \frac{(-i)^n}{n!}\left\{\int d^3y D_{\mu\nu}(y)G^{\mu\nu}\right\}^n \exp-\left\{\frac{1}{2}i\int d^3x d^3y M(x)\Delta(x-y)M(y)\right\}$$

$$\tag{25}$$

with

$$G_{\mu\nu} = \frac{3}{4}\frac{\delta^2}{\delta K^\mu(y)\delta K^\nu(y)} - \frac{1}{4}\frac{\delta^2}{\delta J(y)\delta L^{\mu\nu}(y)}$$
$$-\frac{1}{2}g_{\mu\nu}\left[\frac{1}{2}\frac{\delta^2}{\delta K^\alpha(y)\delta K_\alpha(y)} - m^2\frac{\delta^2}{\delta J(y)\delta J(y)} - \frac{1}{2}g_{\alpha\beta}\frac{\delta^2}{\delta J(y)\delta L_{\alpha\beta}(y)}\right] \tag{26}$$

enables us to repeat the procedure adopted earlier for the 1 + 1 dimensional case in verifying (15).Thus

$$ig^{\alpha\beta}\frac{\delta^3 Z}{\delta D^{\alpha\beta}(u)\delta J(s)\delta J(t)}\bigg|_{J=0=D^{\alpha\beta}}$$
$$=-\frac{i}{2}\begin{bmatrix}(-i\Delta(t-u))\left([\partial^\mu_{(s)}\partial^{(s)}_\mu+m^2](-i\Delta(s-u))\right)+\\(-i\Delta(s-u))\left([\partial^\mu_{(t)}\partial^{(t)}_\mu+m^2](-i\Delta(t-u))\right)+4m^2\left(-i\Delta(t-u)\right)\left(-i\Delta(s-u)\right)\end{bmatrix} \quad (27)$$

$$=\frac{1}{2}i\{i\Delta(t-u)[i\delta^{(3)}(s-u)]+i\Delta(s-u)[i\delta^{(3)}(t-u)]-4m^2(-i\Delta(t-u))(-i\Delta(s-u))\} \quad (28)$$

The l.h.s. of eq.(27) being $i^3 g^{\alpha\beta}\langle 0|T^*\left(\Theta_{\alpha\beta}(u)\phi(s)\phi(t)\right)|0\rangle_c$, eq.(28) gives on using (18b)

$$g^{\alpha\beta}\langle 0|T^*\left(\Theta_{\alpha\beta}(u)\phi(s)\phi(t)\right)|0\rangle_c+\frac{1}{2}i\{\delta^{(3)}(s-u)+\delta^{(3)}(t-u)\}\langle 0|T^*\left(\phi(s)\phi(t)\right)|0\rangle_c$$
$$=\langle 0|T^*\left(\Delta(u)\phi(s)\phi(t)\right)|0\rangle_c \quad (29)$$

which is (5) , remembering eq.(7c) for $\Delta(x)$.

### 4. A check of eqs.(5) and (5a) with boundaries

Let's now first examine the validity of (5a) in the constrained configuration; for this purpose we need to incorporate the Dirichlet boundary condition $\varphi(t, x^1 = -a) = 0$ into the generating functional given by eqs. (18) and (21) above and a simple way to do this is to adopt the method of Bordag et al. [ 5 ] in which one uses the Dirac δ-function to redefine the generating functional in (18) as

$$\exp(iZ[J,K])=\frac{1}{W_0}\int[D\phi]\delta\left(\phi(t,x^1=-a)\right)\exp i\int d^2x\left(L+J\phi+Km^2\phi^2\right) \quad (30)$$

Reworking (30) results in

$$\exp(iZ[J,K])=C\int[D\phi][Db]\exp i\int d^2x\left(L+J\phi+Km^2\phi^2+\delta(x^1+a)b(x)\phi(x)\right) \quad (31)$$

with C a normalizing factor and b(x) an auxiliary field which exists on the plate only; in other words b(x) is a function of the time variable alone. A similar effort on (20) gives

$$\exp iZ\left[J,K^{\alpha},D^{\mu\nu}\right] = C_1\int[D\phi][Db]\exp i\int d^2x\left(L+J\phi+K^{\alpha}\partial_{\alpha}\phi+D_{\mu\nu}\hat{\Theta}_c^{\mu\nu}+\delta\left(x^1+a\right)b(x)\phi(x)\right) \tag{32}$$

The effect of the boundary condition now changes (31) to a modified version of (18) namely,

$$\exp iZ[J,K] = N\sum_0^{\infty}\frac{(-i)^n}{n!}\left(m^2\int d^2xK(x)\frac{\delta^2}{\delta J^2(x)}\right)^n \exp-\frac{1}{2}i\int dxdy J(x)\Delta_s(x-y)J(y) \tag{33}$$

with N a normalizing factor given by

$$N = C\int[Dc]\exp-\frac{1}{2}i\int d^2xd^2y\,\delta\left(x^1+a\right)\delta\left(y^1+a\right)c(x)\Delta(x-y)c(y) \tag{34}$$

c(x) being an auxiliary field defined further below in (37a). In eq.(34) the $\Delta(x-y)$ is identical to that defined by (18a) while the $\Delta_s(x-y)$ used in (33) is defined by

$$\Delta_s(x-y) = \Delta(x-y) - \Delta_0(x-y) \tag{34a}$$

A formal derivation of the second term in $\Delta_s(x-y)$ is given further below ; besides N, eqs. (33) and (18) which define the generating functional for the constrained and unconstrained configurations respectively, differ in the definition of the propagator to be adopted for the calculations below, it being $i\Delta_s(x-y)$ or $i\Delta(x-y)$ for the appropriate case. For eq.(32) one obtains similarly

$$\exp iZ[J,K^{\alpha},D^{\mu\nu}]$$
$$= N_1\sum_0^{\infty}\frac{(-i)^n}{n!}\left(\int d^2xD^{\alpha\beta}H_{\alpha\beta}\right)^n \exp-\frac{1}{2}i\int d^2xd^2y\left(J+K^{\mu}\vec{\partial}_{\mu}\right)_x\Delta_s(x-y)\left(J+\bar{\partial}_{\mu}K^{\mu}\right)_y \tag{35}$$

with $H_{\alpha\beta}$ as in (21a). To make the above derivation reader-friendly, we shall now go through some of the relevant steps leading from (31) to (34); let's first rewrite (31) as

$$\exp(iZ[J,K]) = C\int [D\phi][Db] \left(\frac{(-i)^n}{n!} m^2 \int d^2 x K(x) \frac{\delta^2}{\delta J^2(x)}\right)^n \exp i \int d^2 x \left(L + J\phi + \delta(x^1 + a)b(x)\phi(x)\right)$$
(36)

On transforming to momentum space and changing variables as is done in standard textbooks[7], one obtains besides the familiar $\left(\frac{1}{2}\partial^\mu \phi' \partial_\mu \phi' - \frac{1}{2} m^2 \phi'^2\right)$ due to the shift in the field $\phi(x)$ to $\phi'(x)$, extra terms; in coordinate space the latter are of the form

$$\int_{x^+}\int_{y^+} b(x)\Delta(x-y)b(y) + \int_{x^+}\int_{y} b(x)\Delta(x-y)J(y) + \int_{x}\int_{y^+} J(x)\Delta(x-y)b(y) \quad (37)$$

where $\int_{x^+}$ and $\int_{x}$ are symbols for $\int d^2 x \,\delta(x^1 + a)$ and $\int d^2 x$ respectively. Let's now introduce a new auxiliary field defined by

$$b(x) = c(x) - \int_{u^+}\int_{z} J(z)\Delta(z-u)\Delta^{-1}(u-x) \quad (37a)$$

where

$$\int_{u^+} \Delta^{-1}(z-u)\Delta(u-x) = (2\pi)\delta(z^0 - x^0) \quad (37b)$$

when $z^1$ *and* $x^1$ are restricted to the plate on which $\phi(t, x^1 = -a) = 0$. The reader is referred to the Appendix of this paper for the relevant steps involved in deriving (37b). This definition of $\Delta^{-1}(x-y)$ is an adaptation of eq.(25) in Bordag and Lindig [6] to $1+1$ dimensions. Using eqs.(37a) and (37b) in the first term in (37) leads to

$$\int_{x^+}\int_{y^+} c(x)\Delta(x-y)c(y) - \int_{x^+}\int_{y} c(x)\Delta(x-y)J(y) - \int_{x}\int_{y^+} J(x)\Delta(x-y)c(y) +$$
$$\int_{x}\int_{u^+}\int_{s^+}\int_{y} J(x)\Delta(x-u)\Delta^{-1}(u-s)\Delta(s-y)J(y) \quad (38)$$

Repeating this shift for the second and third terms of (37) easily yields on simplification

$$\int_{x^+}\int_{y^+} c(x)\Delta(x-y)c(y) - \int_z\int_s\int_{u^+}\int_{y^+} J(z)\Delta(z-u)\Delta^{-1}(u-y)\Delta(y-s)J(s) \qquad (39)$$

the first term in (39) being the exponent in the normalization factor N in (34). With the label

$$\Delta_0(x-y) \equiv \int_{u^+}\int_{v^+} \Delta(x-u)\Delta^{-1}(u-v)\Delta(v-y) \qquad (40)$$

and the definition of $\Delta_s(x-y)$ given in (34a) one can now trace the first term in the exponent in (33) to the shift of $\phi(x)$ to $\phi'(x)$, with the second accounted for by (39) and (40). Thus the modification from $\Delta(x-y)$ to $\Delta_s(x-y)$ can be traced to a successive shift in the fields $\phi(x)$ and $b(x)$. We remind the reader here that in the case of a pair of plates with conductor boundary conditions as discussed by Bordag and Lindig [ 6 ] one has beside a pair of Dirac δ - functions in the generating functional for the connected Green's functions a new photon propagator represented by eq.(26) in their paper [6]; eq.(34a) above is therefore the counterpart of eq.(26) of Ref.6.

The above steps also account for the derivation of (35) from (32); let's however anticipate here that it is this addition to the naïve propagator $\Delta(x-y)$ that will be crucial in invalidating the trace identity in 2 + 1 dimensions, while in the 1 + 1 case the trace identity (15) will be maintained because $g_{\alpha\beta}g^{\alpha\beta} = 2$ and because the canonical scale dimension d = 0.

Let's check this out now. From (33) one obtains

$$i\frac{\delta^3 Z[J,K]}{\delta K(u)\delta J(s)\delta J(t)}\bigg|_{J=0=K} \equiv i^3 \langle 0|T^*\left(\Delta(u)\phi(s)\phi(t)\right)|0\rangle_c = -2im^2\left(-i\Delta_s(t-u)\cdot -i\Delta_s(s-u)\right) \qquad (41a)$$

while (35) yields

$$i\frac{\delta^3 Z}{\delta D^{\alpha\beta}(u)\delta J(s)\delta J(t)}\bigg|_{J=0=K=D^{\alpha\beta}} \equiv i^3 \langle 0|T^*\left(\hat{\Theta}_{\alpha\beta}(u)\phi(s)\phi(t)\right)|0\rangle_c$$

$$= -2i\left\{\left(-i\partial_\alpha^{(u)}\Delta_s(t-u)\right)\left(-i\partial_\beta^{(u)}\Delta_s(s-u)\right) - \frac{1}{2}g_{\alpha\beta}\left[\begin{array}{c}\left(-i\partial_\nu^{(u)}\Delta_s(t-u)\right)\left(-i\partial^\nu_{(u)}\Delta_s(s-u)\right)\\ -m^2\left(-i\Delta_s(t-u)\right)\left(-i\Delta_s(s-u)\right)\end{array}\right]\right\} \qquad (41b)$$

Eqs. (41a) and (41b) immediately lead to

$$g_{\alpha\beta}\langle 0|T^*(\hat{\Theta}_c^{\alpha\beta}(u)\phi(s)\phi(t))|0\rangle_c = 2m^2(-i\Delta_s(t-u))(-i\Delta_s(s-u)) = \langle 0|T^*(\Delta(u)\phi(s)\phi(t))|0\rangle_c \quad (42)$$

The first and last equality in (42) thus establishes the validity of (5a) in the constrained configuration adopted in this paper; further, just as in the derivation of eq.(23), the first equality has been got purely on account of the cancellation between the first two terms of the r.h.s. of (41b) because $g_{\alpha\beta}g^{\alpha\beta} = 2$ in 1 + 1 dimensions. In other words, eq.(15) is preserved when one transits from free space to the constrained configuration defined in this paper earlier.

Following the same steps as adopted above for the 1 + 1 dimensional case we shall now take up the verification of (5) in 2 + 1 dimensions. We begin with a suitably modified version of eq.(24) to include the boundary condition $\phi(t, x^1, x^2 = -a) = 0$, namely,

$$\exp iZ[J, K^\mu, D^{\alpha\beta}, L^{\mu\nu}] = C\int[D\phi][Db]\ \exp i\int d^3x \begin{pmatrix} L + J\phi + K^\nu\partial_\nu\phi \\ +L^{\alpha\beta}\partial_\alpha\partial_\beta\phi + D_{\mu\nu}\Theta_M^{\mu\nu} + \delta(x^2 + a)b(x)\phi(x) \end{pmatrix} \quad (43)$$

Reworking (43) in terms of the functional derivative operator $G_{\mu\nu}$ defined in eq.(26) we get

$$\exp iZ[J, K^\alpha, L^{\mu\nu}, D^{\mu\nu}]$$

$$= \sum_0^\infty \frac{(-i)^n}{n!}\left\{\int d^3y D_{\mu\nu}(y)G^{\mu\nu}\right\}^n C$$

$$\left\{\int\int[D\phi][Db]\exp i\int d^3x\left(L + J\phi + K^\mu\partial_\mu\phi(x) + L^{\mu\nu}\partial_\mu\partial_\nu\phi + \delta(x^2 + a)b(x)\phi(x)\right)\right\} \quad (43a)$$

$$= N\sum_0^\infty \frac{(-i)^n}{n!}\left(\int d^3x D_{\mu\nu}G^{\mu\nu}\right)^n \exp -\frac{1}{2}i\int d^3x d^3y M(x)\Delta_s(x-y)M(y) \quad (43b)$$

with M(x) as defined in (24a) and the derivative operators involved in M(x), for example $\partial_x$, acting on the $\Delta(x-y)$. In (43b) we have used the same definition as before (cf. eq.(34)) for N

$$N = C\int [Dc] \exp -\frac{1}{2}i \int_{x^+} \int_{y^+} c(x)\Delta(x-y)c(y) \tag{44}$$

but with $\int_{x^+}$ being a symbol for $\int d^3x \delta(x^2+a)$. Additionally, in the exponent in (43b),

$$\Delta_s(x-y) = \Delta(x-y) - \Delta_0(x-y)$$

with the first term - given by $\Delta(x-y) = \int \frac{d^3p}{(2\pi)^3} \frac{\exp i\, p\cdot(x-y)}{p^2 - m^2 + i\varepsilon}$ - due to the shift of $\phi(x)$ to $\phi'(x)$ so that the Lagrangian density now has the familiar form $\left(\frac{1}{2}\partial^\mu\phi'\partial_\mu\phi' - \frac{1}{2}m^2\phi'^2\right)$, and the second

from the manipulations involved in rewriting the following terms

$$\int_{x^+}\int_{y} b(x)\Delta(x-y)M(y) + \int_{x}\int_{y^+} M(x)\Delta(x-y)b(y) + \int_{x^+}\int_{y^+} b(x)\Delta(x-y)b(y) \tag{45}$$

in terms of a new auxiliary field c(x) defined by

$$b(x) = c(x) - \int_{u^+}\int_z M(z)\Delta(z-u)\Delta^{-1}(u-x) \tag{45a}$$

to obtain

$$\int_{x^+}\int_{y^+} c(x)\Delta(x-y)c(y) - \int_s\int_z\int_{v^+}\int_{y^+} M(z)\Delta(z-y)\Delta^{-1}(y-v)\Delta(v-s)M(s) \tag{46}$$

Notice that eq.(45a) is quite similar to eq. (37a), save that the former involves besides the integration being done in 2 + 1 dimensions, the operator M(x); the second term in (46) defines

$$\Delta_0(x-y) \equiv \int_{v^+}\int_{u^+} \Delta(x-v)\Delta^{-1}(v-u)\Delta(u-y) \tag{46a}$$

with the counterpart of (37b) given by

$$\int_{x^+} \Delta(z-x)\,\Delta^{-1}(x-y) = (2\pi)^2 \delta^{(2)}(z-y) \tag{46b}$$

when $z^2$ and $y^2$ are restricted to the plane defined by the Dirichlet boundary condition $\phi(t, x^1, x^2 = -a) = 0$. While deferring more details on the $\Delta^{-1}(x-y)$ to the Appendix we shall now take up for verification the trace identity given by (5) in the constrained configuration. For convenience let's rewrite below the generating functional

$$\exp iZ[J, K^\alpha, L^{\mu\nu}, D^{\mu\nu}] = N \sum_0^\infty \frac{(-i)^n}{n!} \left( \int d^3 x D_{\mu\nu} G^{\mu\nu} \right)^n \exp -\frac{1}{2} i \int d^3 x d^3 y M(x) \Delta_s(x-y) M(y) \quad (47)$$

From (47) one gets

$$i \frac{\delta^2 Z}{\delta J(s) \delta J(t)} \bigg|_{sources=0} = \left( -i\Delta(t-s) + i\Delta_0(t-s) \right) = i^2 \left\langle 0 \left| T^* \left( \phi(s)\phi(t) \right) \right| 0 \right\rangle_{c1} \quad (47a)$$

where an additional subscript 1 is included in the connected Green's function in (47a) to distinguish it from the free-space version in the 2 + 1 –dimensions displayed in (18b) and

$$ig^{\mu\nu} \frac{\delta^3 Z}{\delta D^{\mu\nu} \delta J(s) \delta J(t)} \bigg|_{J=0=D^{\mu\nu}} = i^3 g^{\mu\nu} \left\langle 0 \left| T^* \left( \Theta_{\mu\nu}(u)\phi(s)\phi(t) \right) \right| 0 \right\rangle_c$$

$$= (-i)(\frac{1}{2}) \{ \left( -i\Delta(s-u) + i\Delta_0(s-u) \right) \left( -i\partial_\alpha^{(u)} \partial_{(u)}^\alpha \Delta(u-t) + i\partial_\alpha^{(u)} \partial_{(u)}^\alpha \Delta_0(u-t) \right)$$

$$+ \left( -i\Delta(t-u) + i\Delta_0(t-u) \right) \left( -i\partial_\alpha^{(u)} \partial_{(u)}^\alpha \Delta(u-s) + i\partial_\alpha^{(u)} \partial_{(u)}^\alpha \Delta_0(u-s) \right) \}$$

$$+ (-i)(\frac{1}{2} m^2) g_{\alpha\beta} g^{\alpha\beta} 2 \{ \left( -i\Delta(t-u) + i\Delta_0(t-u) \right) \left( -i\Delta(s-u) + i\Delta_0(s-u) \right) \} \quad (48)$$

Remembering that $g_{\alpha\beta} g^{\alpha\beta} = 3$ one can now combine two terms in (48) as follows:

a.  $(-i)(\frac{1}{2}) \{ \left( -i\Delta(s-u) + i\Delta_0(s-u) \right) \left( -i\partial_\alpha^{(u)} \partial_{(u)}^\alpha \Delta(u-t) + i\partial_\alpha^{(u)} \partial_{(u)}^\alpha \Delta_0(u-t) \right) \}$

$+ (-i)(\frac{1}{2}) \{ m^2 \left( -i\Delta(u-s) + i\Delta_0(u-s) \right) \left( -i\Delta(u-t) + i\Delta_0(u-t) \right) \}$

$= (-i)(\frac{1}{2}) \{ \left( -i\Delta(s-u) + i\Delta_0(s-u) \right) \left( i\delta^{(3)}(u-t) + i\partial_{(u)}^\mu \partial_\mu^{(u)} \Delta_0(u-t) + m^2 i\Delta_0(u-t) \right) \quad (49a)$

b. $(-i)(\frac{1}{2})\{(-i\Delta(u-t)+i\Delta_0(u-t))(-i\partial_\alpha^{(u)}\partial_{(u)}^\alpha \Delta(u-s)+i\partial_\alpha^{(u)}\partial_{(u)}^\alpha \Delta_0(u-s))\}$

$+(-i)(\frac{1}{2})\{m^2(-i\Delta(u-t)+i\Delta_0(u-t))(-i\Delta(u-s)+i\Delta_0(u-s))\}$

$= (-i)(\frac{1}{2})\{(-i\Delta(u-t)+i\Delta_0(u-t))(i\delta^{(3)}(u-s)+i\partial_{(u)}^\mu \partial_\mu^{(u)}\Delta_0(u-s)+m^2 i\Delta_0(u-s))\}$ (49b)

The remainder term in (48) is now

$$(-i)\{2m^2(-i\Delta(u-t)+i\Delta_0(u-t))(-i\Delta(u-s)+i\Delta_0(u-s))\} \qquad (49c)$$

From the generating functional

$$\exp iZ[J,K] = \frac{1}{W_0}\int [D\phi][Db]\exp i\int d^3x\left(L+J\phi+Km^2\phi^2+\delta(x^2+a)b(x)\phi(x)\right)$$

$$= \frac{1}{W_0}\sum_0^\infty \frac{(-i)^n}{n!}\left(\int d^3x K(x)m^2 \frac{\delta^2}{\delta J^2(x)}\right)^n \int [D\phi][Db]\exp i\int d^3x\left(L+J\phi+\delta(x^2+a)b(x)\phi(x)\right)$$

(50)

one can now show that

$$i\frac{\delta^3 Z}{\delta K(u)\delta J(s)\delta J(t)}\bigg|_{J=0=K} \equiv i^3\langle 0|T^*(\Delta(u)\phi(s)\phi(t))|0\rangle_c$$
$$= -2im^2(-i\Delta(u-t)+\Delta_0(u-t))(-i\Delta(u-s)+\Delta_0(u-s))$$

(51)

Clearly eq.(51) is the remainder term given by (49c); from (48) and (51) we thus obtain

$i^3 g^{\mu\nu}\langle 0|T^*(\Theta_{\mu\nu}(u)\phi(s)\phi(t))|0\rangle_c - i^3\langle 0|T^*(\Delta(u)\phi(s)\phi(t))|0\rangle_c$

$=(-i)(\frac{1}{2})\{(-i\Delta(s-u)+i\Delta_0(s-u))i\delta^{(3)}(u-t)+(-i\Delta(u-t)+i\Delta_0(u-t))i\delta^{(3)}(u-s)\}+A$

Let's now use eq.(47a) to rewrite the r.h.s. above as

$$= (-i)(\frac{1}{2})\{i^2 \langle 0|T^*(\phi(s)\phi(u))|0\rangle_{c1} i\delta^{(3)}(u-t) + i^2 \langle 0|T^*(\phi(u)\phi(t))|0\rangle_{c1} i\delta^{(3)}(u-s)\} + A$$

(52)

with

$$A = (-i)(\frac{1}{2})\{(-i\Delta(s-u) + i\Delta_0(s-u))\left(i\partial^\mu_{(u)}\partial^{(u)}_\mu \Delta_0(u-t) + m^2 i\Delta_0(u-t)\right) +$$

$$(-i\Delta(t-u) + i\Delta_0(t-u))\left(i\partial^{(u)}_\alpha \partial^\alpha_{(u)}\Delta_0(u-s) + m^2 i\Delta_0(u-s)\right)\}$$

(53)

We shall now rework eq.(52) as

$$g^{\mu\nu}\langle 0|T^*(\Theta_{\mu\nu}(u)\phi(s)\phi(t))|0\rangle_c + \frac{1}{2}\langle 0|T^*(\phi(s)\phi(t))|0\rangle_{c1}\{i\delta^{(3)}(u-t) + i\delta^{(3)}(u-s)\}$$

$$= \langle 0|T^*(\Delta(u)\phi(t)\phi(s))|0\rangle_c + iA$$

(54)

On comparing eq.(54) with (29) above, we notice immediately that the nonzero A in (53) now invalidates the trace identity (29) that was otherwise preserved in free space in $2+1$ dimensions. It is instructive to have a second look at A; for this purpose we recall from eq.(40) that

$$\Delta_0(x-y) \equiv \int_{v^+}\int_{u^+} \Delta(x-v)\Delta^{-1}(v-u)\Delta(u-y)$$

Clearly therefore

$$i\left(\partial^{(x)}_\mu \partial^\mu_{(x)} + m^2\right)\Delta_0(x-y) = -i\int_{z^+}\int_{s^+} \delta^{(3)}(x-z)\Delta^{-1}(z-s)\Delta(s-y)$$

$$= -i\int\int_{s^+} d^3z\, \delta(z^2+a)\delta^{(3)}(x-z)\Delta^{-1}(z-s)\Delta(s-y) \quad (55)$$

On doing the z- integration in (55) we will get

$$i\left(\partial_\mu^{(x)}\partial^\mu_{(x)} + m^2\right)\Delta_0(x-y) = -i\delta(x^2+a)\int_{s^+} \Delta^{-1}(x-s)\Delta(s-y) \tag{55a}$$

By virtue of the Dirac $\delta$ - function on the r.h.s. of (55a) it is clear that the integral will yield $\delta^{(2)}(x-y)$ when $y^2$ is also restricted to the plate on which the 2 + 1 dimensional version of eq.(37b),viz.

$$\int_{u^+} \Delta^{-1}(z-u)\Delta(u-x) = \delta^{(2)}(z-x) \tag{55b}$$

holds. While referring the reader to the Appendix below for further details on the derivation of both eqs.(37b) and (55b),we shall now use eqs.(55) and (55a) in (53) to obtain

$$A = (-i)(\frac{1}{2})\{(-i\Delta(s-u) + i\Delta_0(s-u))\,\{-i\delta(u^2+a)\int_{x^+} \Delta^{-1}(u-x)\Delta(x-t)\} +$$

$$(-i\Delta(t-u) + i\Delta_0(t-u))\,\{-i\delta(u^2+a)\int_{x^+} \Delta^{-1}(u-x)\Delta(x-s)\}\,\} \tag{56}$$

Note that when $t^2$ and $s^2$ are set to - a one can now invoke (55b) to rework A as

$$A = (-i)(\frac{1}{2})\{(-i\Delta(s-u) + i\Delta_0(s-u))\,\{-i\delta(u^2+a)\delta^{(2)}(u-t)\}$$

$$+(-i\Delta(t-u) + i\Delta_0(t-u))\,\{-i\delta(u^2+a)\delta^{(2)}(u-s)\}\} \tag{57}$$

$$=(-i)(\frac{1}{2})\{\,i^2\left\langle 0\left|T^*\left(\phi(s)\phi(u)\right)\right|0\right\rangle_{c1} \{-i\delta(u^2+a)\delta^{(2)}(u-t)\} + i^2\left\langle 0\left|T^*\left(\phi(u)\phi(t)\right)\right|0\right\rangle_{c1}$$

$$\{-i\delta(u^2+a)\delta^{(2)}(u-s)\}\}$$

Thus

$$iA = -\left(\frac{1}{2}\right)\left[\left\langle 0\left|T^*\left(\phi(s)\phi(u)\right)\right|0\right\rangle_{c1} \{-i\delta(u^2+a)\delta^{(2)}(u-t)\}\right.$$

$$+ \left\langle 0 \left| T^* \left( \phi(u)\phi(t) \right) \right| 0 \right\rangle_{c1} \{-i\delta(u^2 + a)\delta^{(2)}(u-s)\} \Big] \tag{58}$$

Identifying $t^2$ and $s^2$ with $-a$ enables us to rewrite eq.(58) as

$$iA = -\left(\frac{1}{2}\right) \Big[ \left\langle 0 \left| T^* \left( \phi(s)\phi(u) \right) \right| 0 \right\rangle_{c1} \{-i\delta(u^2 - t^2)\delta^{(2)}(u-t)\}$$

$$+ \left\langle 0 \left| T^* \left( \phi(u)\phi(t) \right) \right| 0 \right\rangle_{c1} \{-i\delta(u^2 - s^2)\delta^{(2)}(u-s)\} \Big] \tag{59}$$

$$= -\left(\frac{1}{2}\right) \left\langle 0 \left| T^* \left( \phi(s)\phi(t) \right) \right| 0 \right\rangle_{c1} \{-i\delta(u^2 - t^2)\delta^{(2)}(u-t) - i\delta(u^2 - s^2)\delta^{(2)}(u-s)\}$$

$$\tag{60}$$

Clearly eq.(60) matches exactly with the second term on the l.h.s. of eq.(54) strongly suggesting that in the constrained configuration that we have adopted for the purposes of this paper there is an *induced* anomalous dimension of (1/2) which is equal to the canonical scale dimension for the massive real scalar field; as emphasized earlier this is peculiar to the planar case and does not obtain for the 1 + 1 dimensional case for which the canonical scale dimension of $\phi(x)$ is zero. Let's also point out now that the term 'induced' has been used here *intentionally* in the sense that it has appeared because of the introduction of the boundary, implying thereby that it would disappear when the boundary is removed. In the sense used in this paper therefore there is a qualitative difference between 1 + 1 and 2 + 1 dimensions in so far as the maintenance of the familiar trace identity is concerned on a boundary.

**Discussion:**

To trace the origin of this 'anomaly' - if one can call it so - in the trace identity it is useful to recall here the functional derivative operator in (26), namely,

$$G_{\mu\nu} = \frac{3}{4}\frac{\delta^2}{\delta K^\mu(y)\delta K^\nu(y)} - \frac{1}{4}\frac{\delta^2}{\delta J(y)\delta L^{\mu\nu}(y)}$$
$$-\frac{1}{2}g_{\mu\nu}\left[\frac{1}{2}\frac{\delta^2}{\delta K^\alpha(y)\delta K_\alpha(y)} - m^2\frac{\delta^2}{\delta J(y)\delta J(y)} - \frac{1}{2}g_{\alpha\beta}\frac{\delta^2}{\delta J(y)\delta L_{\alpha\beta}(y)}\right]$$

Notice that on doing the functional differentiation to calculate the l.h.s. of eq.(48) the first and third term in $g^{\mu\nu}G_{\mu\nu}$ cancel exactly when $g^{\mu\nu}g_{\mu\nu} = 3$; this is the reason why the r.h.s. of (48) is slimmer than expected in that it contains only the contributions from the three remaining terms in $G_{\mu\nu}$. This happens in 1 + 1 dimensions too as a quick glance at eq.(22) confirms, but a second look at eqs.(21a) versus (26) shows that there are two extra terms in $G_{\mu\nu}$, namely the second and the fifth, both due to the nonzero field virial in the planar - as opposed to the 1 + 1 dimensional - case. To take the argument further, the effect of these extra terms shows up in the constrained case due to the additional presence of the $\Delta_0(x-y)$ in the modified propagator $\Delta_s(x-y)$ - see (47a) above - resulting from the introduction of a boundary in the 2 + 1 dimensional case. Put differently, the twin effects of the nonzero virial and the modified propagator $\Delta_s(x-y)$ are together responsible for the failure of the trace identity in the 2 + 1 dimensional case when one transits from free space to a constrained one, the latter distinguished by a boundary on which the Dirichlet condition $\varphi(t, x^1, x^2 = -a) = 0$ is imposed.

It is important here to underline an earlier observation in this section that the coefficient of the anomalous term – see eq.(60) above – is the same as the canonical scale dimension which is ½ in the 2 + 1 dimensional case; the absence of the 'anomaly' in 1 + 1 dimensions can thus be regarded as due to the zero value of the canonical scale dimension for $\phi(x)$ in this case. It also stands to reason that the 'anomalous' result reported in this paper for the trace identity associated with the Lagrangian for the noninteracting massive real scalar field in 2 + 1 dimensions will definitely prevail – given the premises used in the calculation here - when one extends this investigation to other field theories especially those where the canonical scale dimension is not zero; this is in progress and will be reported elsewhere.

Let's now conclude this paper on a hypothetical but hopeful note; in doing so we are encouraged by the fact that Casimir forces have now been measured precisely, the experiment by Lamoreaux [8] being the first in a modern phase of Casimir force measurements. While relegating the details of that experiment and subsequent developments therein to the review by Bordag et al.[4] we must point out here that Casimir forces also play a significant role in micro-and nanometer size structures. More precisely, it is now recognized [9] that these forces need to be factored in the design and modeling of MEMS and nanoelectromechanical systems(NEMS); as a spin-off from these experimental advances we hazard the guess that the theoretical view-point taken in two recent papers by Cavalcanti[10] and Hertzberg et al. [11] on the Casimir piston could be of some relevance to the results of this paper.

In other words, the idea we are putting across now is at best tentative and needs to be worked on carefully, more so as the Dirichlet boundary condition is now imposed on a single plate in this paper and not on the walls of a two-dimensional box of dimensions L x b as in Ref.10.However, to pursue this example further, one would now have to calculate the energy-density on the movable piston of Ref.10 for both finite L - with the Dirichlet boundary condition on the single plate which now becomes the wall – and then let $L \to \infty$ thus inferring the force on the piston. We should then be able to estimate the role of the piston-base separation on the Casimir force and relate our result to that obtained by Hertzberg et al.[11];additional features involving aspect ratios as discussed in earlier papers[12] could also perhaps be relevant given the role of dilatations in this paper.

Whereas the above remarks anticipate a future course for the work done in this paper in a vague but qualitative fashion while tuning it to the recent laboratory work mentioned above on the Casimir effect, let's comment finally that an extension of the work reported here to the Maxwell-Chern-Simon Lagrangian in 2 + 1 dimensions is presently in progress; the details of which will be published elsewhere.

# Appendix

We present here the relevant formulae for $\Delta^{-1}(x-y)$ beginning with $1+1$ dimensions; the propagator that we shall use is given by

$$i\Delta(x-y) = i\int \frac{d^2p}{(2\pi)^2} \frac{\exp i\,p\cdot(x-y)}{p^2 - m^2 + i\varepsilon} \tag{A1}$$

Performing a Wick rotation gets us

$$i\Delta(x-y) = -i^2 \int_{-\infty}^{\infty} \frac{dp_0}{4\pi} e^{-ip_0(x_0-y_0)} \frac{e^{-\left(p_0^2+m^2\right)^{1/2}|x_1-y_1|}}{\left(p_0^2+m^2\right)^{1/2}} \tag{A2}$$

Undoing the Wick rotation now gives

$$i\Delta(x-y) = -i^2 \int_{-\infty}^{\infty} \frac{dp_0}{4\pi} e^{-ip_0(x_0-y_0)} \frac{e^{-i\left(p_0^2-m^2+i\varepsilon\right)^{1/2}|x_1-y_1|}}{\left(p_0^2-m^2+i\varepsilon\right)^{1/2}} \tag{A3}$$

We shall now define

$$i\Delta^{-1}(x-y) = i\left(-\frac{2}{i}\right)\int_{-\infty}^{\infty} \frac{dp_0}{2\pi} e^{-ip_0(x_0-y_0)} \left(p_0^2 - m^2 + i\varepsilon\right)^{1/2} e^{i\left(p_0^2-m^2+i\varepsilon\right)^{1/2}|x_1-y_1|} \tag{A4}$$

sothat

$$\int_{y^+} i\Delta(x-y)i\Delta^{-1}(y-z) = i^2\left(-\frac{i}{2}\right)\left(-\frac{2}{i}\right)\int_{y^+}\int_{-\infty}^{\infty}\int_{-\infty}^{\infty} \frac{dp_0}{2\pi}\frac{dq_0}{2\pi} e^{-ip_0(x_0-y_0)} e^{-iq_0(y_0-z_0)} \{\frac{e^{-i\left(p_0^2-m^2+i\varepsilon\right)^{1/2}|x_1-y_1|}}{\left(p_0^2-m^2+i\varepsilon\right)^{1/2}}$$

$$\left(q_0^2 - m^2 + i\varepsilon\right)^{1/2} e^{i\left(q_0^2-m^2+i\varepsilon\right)^{1/2}|y_1-z_1|}\}$$

$$\tag{A5}$$

Integrating now over $y^0$ gives $(2\pi)\delta(p_0 - q_0)$ so that (A5) now becomes

$$\int_{y^+} i\Delta(x-y)i\Delta^{-1}(y-z) = i^2 \int_{-\infty}^{\infty} \frac{dp_0}{2\pi} e^{-ip_0(x_0-z_0)} e^{i(p_0^2-m^2+i\varepsilon)^{1/2}|y_1-z_1|} e^{-i(p_0^2-m^2+i\varepsilon)^{1/2}|x_1-y_1|} \tag{A6}$$

with $y^1 = -a$. When $x^1$ and $z^1$ are restricted to the plate so that $x^1 = -a = z^1$ one easily gets $\delta(p_0 - q_0)$ for the integral; this is (37b).

Naturally, an extension of the above exercise to 2 + 1 dimensions is immediate and begins with

$$i\Delta(x-y) = i\int \frac{d^3p}{(2\pi)^3} \frac{\exp i p \cdot (x-y)}{p^2 - m^2 + i\varepsilon} \tag{A7}$$

Integrating over $p_2$ and repeating the same steps as was done in working through (A1) to (A3) one obtains

$$i\Delta(x-y) = -i^2 \int \frac{d^2p}{2(2\pi)^2} e^{-ip\cdot(x-y)} \frac{e^{-i(p^2-m^2+i\varepsilon)^{1/2}|x_2-y_2|}}{(p^2-m^2+i\varepsilon)^{1/2}} \tag{A8}$$

One now defines

$$i\Delta^{-1}(x-y) = i\left(-\frac{2}{i}\right)\int \frac{d^2p}{(2\pi)^2} e^{-ip\cdot(x-y)} (p^2-m^2+i\varepsilon)^{1/2} e^{i(p^2-m^2+i\varepsilon)^{1/2}|x_2-y_2|} \tag{A9}$$

so that the counterpart of (A5) will now be

$$\int_{y^+} i\Delta(x-y)i\Delta^{-1}(y-z) = i^2\left(-\frac{i}{2}\right)\left(-\frac{2}{i}\right)\int_{y^+}\int_{-\infty}^{\infty}\int_{-\infty}^{\infty}\frac{d^2p}{(2\pi)^2}\frac{d^2q}{(2\pi)^2}e^{-ip\cdot(x-y)}e^{-iq\cdot(y-z)}$$

$$\{\frac{e^{-i(p^2-m^2+i\varepsilon)^{1/2}|x_2-y_2|}}{(p^2-m^2+i\varepsilon)^{1/2}}(q^2-m^2+i\varepsilon)^{1/2}e^{i(q^2-m^2+i\varepsilon)^{1/2}|y_2-z_2|}\}$$

(A10)

Remembering that $\int_{x^+}$ is a symbol for $\int d^3x\delta(x^2+a)$, one obtains on doing the space integration and the resulting integration over one of the momentum variables the relation

$$\int_{y^+} i\Delta(x-y)i\Delta^{-1}(y-z) = i^2\int\frac{d^2p}{(2\pi)^2}e^{-ip\cdot(x-z)}e^{-i(p^2-m^2+i\varepsilon)^{1/2}|x_2-y_2|}e^{i(p^2-m^2+i\varepsilon)^{1/2}|y_2-z_2|}$$

(A11)

Note that on identifying $x^2$ and $z^2$ with $-a$ one recovers $\delta^{(2)}(p-q)$ thus yielding the 2-dimensional version of (37b) namely,

$$\int_{y^+} i\Delta(x-y)i\Delta^{-1}(y-z) = i^2\delta^{(2)}(x-z)$$

(A12)

Both eqs.(37b) and (A12) above are the counterparts of eq.(25) of Bordag and Lindig [ 6 ] .